\documentclass[12pt]{article}

\usepackage[margin=1in]{geometry}

\usepackage{graphicx} 
\usepackage[utf8]{inputenc} 
\usepackage[T1]{fontenc}    
\usepackage[hidelinks, colorlinks=true,citecolor=blue]{hyperref} 
\usepackage{url}
\usepackage{booktabs}
\usepackage{amsfonts}
\usepackage{nicefrac}
\usepackage{microtype}
\usepackage{xcolor}
\usepackage[normalem]{ulem}
\usepackage{amsmath}
\usepackage{amssymb}
\usepackage{color}
\usepackage[numbers]{natbib}
\usepackage{caption}
\usepackage{subcaption}
\usepackage{bm}
\usepackage{bbm}
\usepackage[flushleft]{threeparttable}
\usepackage[shortlabels]{enumitem}
\usepackage{wrapfig}
\setlist{leftmargin=6mm}
\usepackage{multirow}
\usepackage{titlesec}
\usepackage{setspace}
\usepackage{array}

\usepackage{mathtools, mathrsfs}
\usepackage{textcomp, gensymb}
\usepackage{dsfont}
\usepackage{soul}

\usepackage{algorithm}
\usepackage{algorithmic}
\newlength\myindent
\usepackage{amsthm}

\newtheorem{conjecture*}{Conjecture}
\theoremstyle{remark}

\allowdisplaybreaks

\title{Atlanta Gun Violence Modeling via \\ Nonstationary Spatio-temporal Point Processes}

\author{
  Zheng Dong, Yao Xie\thanks{Email: yao.xie@isye.gatech.edu}
}

 \date{\small
     H. Milton Stewart School of Industrial \& Systems Engineering \\Georgia Institute of Technology
 }

\begin{document}

\maketitle

 \begin{abstract}
     Analysis of gun violence in the United States has utilized various models based on spatiotemporal point processes. Previous studies have identified a contagion effect in gun violence, characterized by bursts of diffusion across urban environments, which can be effectively represented using the self-excitatory spatiotemporal Hawkes process. The Hawkes process and its variants have been successful in modeling self-excitatory events, including earthquakes, disease outbreaks, financial market movements, neural activity, and the viral spread of memes on social networks. However, existing Hawkes models applied to gun violence often rely on simplistic stationary kernels, which fail to account for the complex, non-homogeneous spread of influence and impact over space and time. To address this limitation, we adopt a non-stationary spatiotemporal point process model that incorporates a neural network-based kernel to better represent the varied correlations among events of gun violence. Our study analyzes a comprehensive dataset of approximately 16,000 gunshot events in the Atlanta metropolitan area from 2021 to 2023. The cornerstone of our approach is the innovative non-stationary kernel, designed to enhance the model's expressiveness while preserving its interpretability. This approach not only demonstrates strong predictive performance but also provides insights into the spatiotemporal dynamics of gun violence and its propagation within urban settings.
 \end{abstract}

\section{Introduction}

Gun violence remains a critical issue in many U.S. cities, influenced by a complex interplay of social and economic factors \citep{braga2010concentration}. Understanding the patterns and determinants of gun violence is essential for local policymakers to develop and implement effective intervention strategies. Previous research on gun-related incidents \citep{fagan2007social, green2017modeling, loeffler2018gun, papachristos2009murder, rosenfeld1999facilitating, short2014gang} has consistently identified a contagious pattern of gun violence, suggesting that these events are not isolated incidents but are interconnected through underlying social and spatial factors. These patterns present a predictive potential that can be leveraged to mitigate future risks of gun violence. Additionally, the availability of rich, real-time data sources in metropolitan areas, such as 911 call records, provides a valuable opportunity for analyzing the underlying dynamics of gun-related events and forecasting the risk of gun violence in urban settings.

\vspace{.1in}
However, modeling and predicting gun violence continues to present significant challenges. A primary obstacle lies in the vast volume and complexity of the data involved. For instance, our dataset comprises nearly 16,000 reported gunshot events in the city of Atlanta, creating a substantial computational burden and necessitating sophisticated analytical methods. Furthermore, the data is multi-modal, as it integrates direct reports from 911 calls with detailed demographic and socioeconomic census data. Previous studies have addressed the spatio-temporal clustering of gun violence \citep{loeffler2018gun, taylor2013contagion} and other types of crime \citep{cohen1999diffusion, mohler2011self, morenoff2001neighborhood, johnson2008repeat, short2010dissipation}, but they often employ simpler models that cannot fully capture the heterogeneous spatio-temporal dynamics of violence spread across different urban contexts. This gap underscores the need for a more tailored framework capable of effectively characterizing these complex event patterns.

\vspace{.1in}
In this paper, we present a point process model designed to address the challenges of modeling and capturing the complex spatio-temporal dynamics underlying gunshot events, and showcase using gunshots data in Atlanta in the United States. Drawing inspiration from previous non-stationary point process models for earthquakes and pandemics \citep{dong2023non, zhu2021imitation}, our approach incorporates a non-stationary influence kernel to represent the intricate contagious effects of gunshot events. Moreover, our model extends beyond existing frameworks by integrating a heterogeneous background intensity, which accounts for the inherent variability in gunshot events influenced by diverse census factors across different geographic regions. Through validation using real-world data, we demonstrate the effectiveness and predictive power of our model compared with the state-of-the-art. The proposed model can be used to understand gun violence propagation, and also enhances our capacity to predict future hotspots and trends, thereby providing actionable insights for law enforcement and public policymakers to prevent urban gun violence.

\vspace{.1in}
The rest of the paper is organized as follows. In the remainder of this section, we continue to review related works. 
Section~\ref{sec:data} introduces the gunshot data collected by the Atlanta Police Department and the public census data used in our study. Section~\ref{sec:model} presents the point process model; Section ~\ref{sec:model-fitting} explains modeling fitting strategies.  Section~\ref{sec:results} presents numerical results on the real data sets to demonstrate the effectiveness and interpretability of our model. Finally, Section~\ref{discussion} concludes the paper with discussions.

\paragraph{Related works.} 

The contagious effect of crime has been previously studied since several early works \citep{fagan2007social, loftin1986assaultive, roth1993understanding}, where researchers modeled gun-related homicides as a point process with contagion characteristics. Crime contagion refers to the phenomenon where the occurrence of violence increases the likelihood of subsequent violence. Notably, most researchers have focused on this contagion at the event level, examining its effects over shorter time scales. They emphasize the contribution of individual events to the heightened crime risk in the following days and in nearby neighborhoods \citep{skogan2008evaluation, webster2013effects}, which is part of the broader spatio-temporal study of crime properties.

\vspace{.1in}
Spatio-temporal clustering patterns of crime events have been documented across various crime types, including violent crime \citep{braga2010concentration, cohen1999diffusion, morenoff2001neighborhood, ratcliffe2008near} and property crime \citep{johnson2008repeat, mohler2011self,short2010dissipation}. Early work by Messner et al. \citep{messner1999spatial} identified a non-random clustering pattern of homicides at the county level, while Ratcliffe and Rengert \citep{ratcliffe2008near} observed similar non-random spatiotemporal clustering at the block level. In contrast to these studies, our research focuses on analyzing violent crime patterns at the event level, without spatial aggregation of crime data. A recent study \citep{loeffler2018gun} highlights the difference between the clustering of gun violence with and without diffusion/contagion effects, demonstrating the existence of such a spatio-temporal diffusion pattern of gun violence in a metropolitan area.

\vspace{.1in}
The application of self-exciting point processes in crime modeling, originally motivated by their use in modeling earthquake occurrences in seismology \citep{Ogata1988}, has been extensively explored to characterize the spatio-temporal clustering patterns of crime events \citep{lewis2012self, mohler2013modeling, mohler2011self, reinhart2018review, zhu2022spatiotemporal, zhuang2019semiparametric}. Early attempts demonstrated the effectiveness of point processes in modeling crime using residential burglary data in Los Angeles \citep{mohler2011self} and civilian death reports in Iraq \citep{lewis2012self}. Subsequent approaches \citep{mohler2014marked, reinhart2018self, zhu2022spatiotemporal} improved point process models for crime by incorporating the events' type, location, and textual information to capture complex crime patterns in various modeling tasks. Our approach extends the commonly used stationary influence kernel to a non-stationary one with a neural network representation, enhancing the model's ability to capture the spatial heterogeneity observed in the occurrences of gunshot events. The effectiveness of our non-stationary kernel is both theoretically and empirically supported by previous studies on deep non-stationary influence kernels \citep{dong2023spatio, dong2023deep, dong2023non, zhu2021imitation} and the numerical results presented in this study.

\vspace{.1in}
An increasing level of gun violence has been observed in many U.S. cities, drawing significant attention from the criminology community. Previous studies \citep{fagan2007social, loeffler2018gun, taylor2013contagion} have suggested the presence of gun violence contagion, which can be driven by various factors, such as the rapid diffusion of human behaviors \citep{de1903laws} or the structure of social networks \citep{green2017modeling, papachristos2009murder, rosenfeld1999facilitating, short2014gang}. Another line of research highlights a strong relationship between gun violence and the characteristics of local communities, including socioeconomic factors \citep{braga2010concentration, kravitz2022inequities, morenoff2001neighborhood, papachristos2009murder} and physical infrastructure \citep{xu2017shooting}. The motivation of this work is that by considering both the contagion of gunshot events and the influence of various exogenous spatial factors across different communities on gun violence risk, we can provide a comprehensive depiction of the dynamics of gun violence in the city of Atlanta.

\section{Atlanta gunshot data}
\label{sec:data}

The gunshot event dataset used in this study was collected by the Atlanta Police Department (APD) in the city of Atlanta, the capital and most populous city in Georgia, and the core of the sixth-largest U.S. metropolitan area. To foster a manageable and supportive living environment, the city is divided into 242 distinct neighborhoods, including notable areas such as Midtown, Downtown, and Buckhead. In this study, we base our data analysis and model development on this neighborhood division, enabling us to understand the influence of local factors on gunshot patterns and to support the development of tailored strategies for enhancing community safety.

\begin{figure}[!t]
\centering
\begin{subfigure}[h]{0.95\linewidth}
\includegraphics[width=\linewidth]{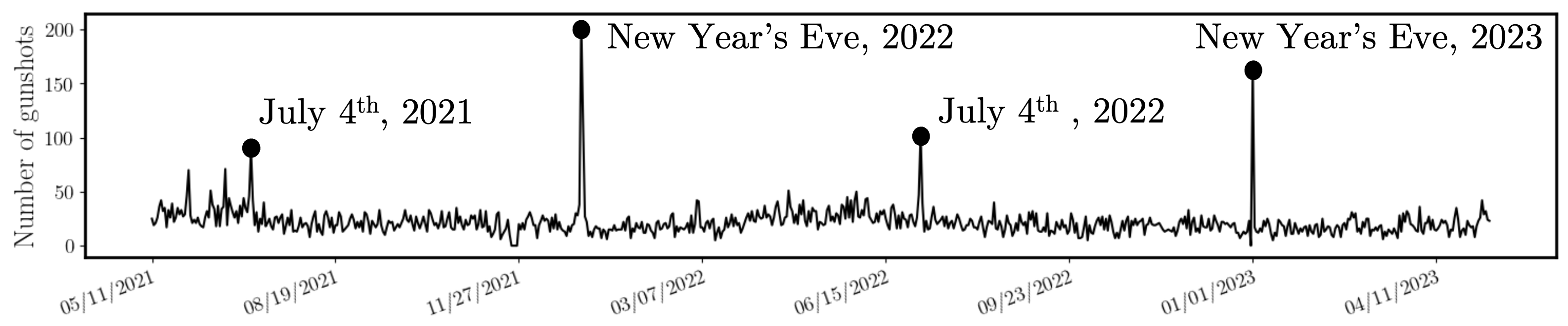}
\end{subfigure}
\caption{Daily number of gunshot events from May 11, 2021, to May 10, 2023, with four peaks of gunshot events highlighted in the line chart.}
\label{fig:gunshot-over-time}
\end{figure}

\begin{figure}[!t]
\centering
\begin{subfigure}[h]{0.4\linewidth}
\includegraphics[width=\linewidth]{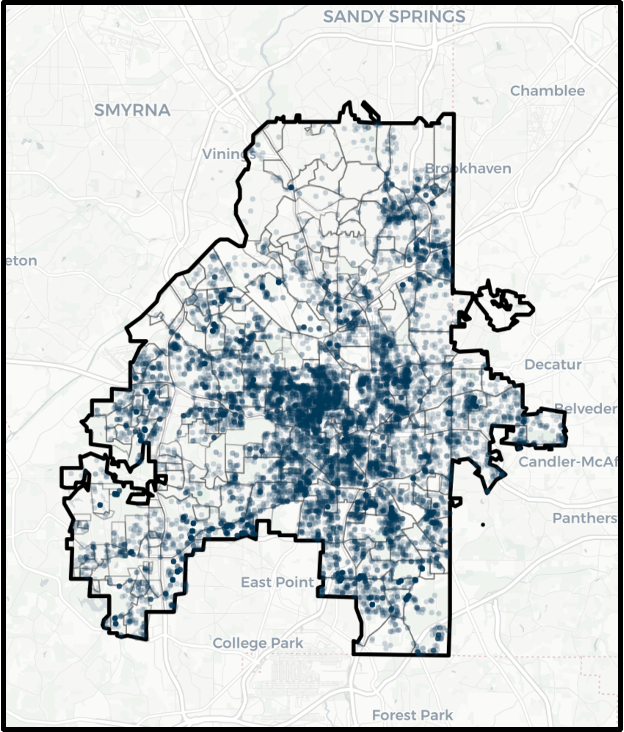}
\caption{All gunshot events}
\end{subfigure}
\begin{subfigure}[h]{0.392\linewidth}
\includegraphics[width=\linewidth]{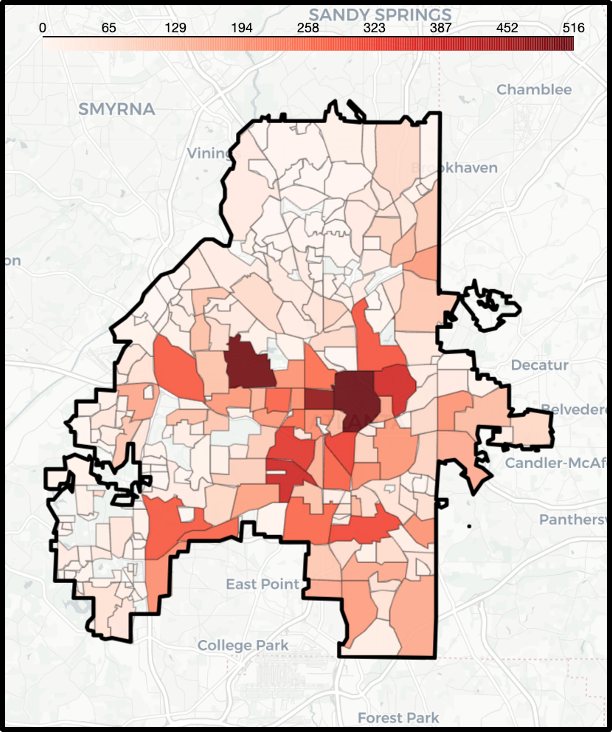}
\caption{Aggregated by neighborhood}
\end{subfigure}
\caption{Distribution of all gunshot events reported in the Atlanta city area from May 11, 2021, to May 10, 2023. The grey lines indicate the boundaries of different neighborhoods. (a) Event-level distribution: Each dot represents a reported gunshot event, with darker areas indicating overlapping dots. (b) Neighborhood-level distribution: The total number of gunshot events in each neighborhood is visualized, with darker colors indicating higher numbers of gunshot events.}
\label{fig:gunshot-distribution}
\end{figure}

\vspace{.1in}

The gunshot dataset comprises a total of 15,940 reported gunshot incidents from May 11, 2021, to May 10, 2023. Each incident's time (accurate to the second) and geographic location (measured in longitude and latitude) are recorded. Figure~\ref{fig:gunshot-over-time} illustrates the daily number of gunshot events over this two-year period. Although the number of gunshot events remains relatively stable over time, there is a noticeable surge in intensity around major holidays. Figure~\ref{fig:gunshot-distribution} visualizes the spatial distribution of all reported gunshot events within the Atlanta city area, both at the event level and neighborhood level. The spatial heterogeneity of gunshot events is evident, with the center (\textit{e.g.}, Downtown) and southern parts of Atlanta experiencing more reported incidents than the northern part of the city (\textit{e.g.}, Buckhead). This non-stationarity in both temporal and spatial dimensions highlights the need for a non-stationary kernel and location-dependent background intensity in the model to account for the potential effects of significant spatial covariates.

\begin{figure}
\centering
\begin{subfigure}[h]{0.4\linewidth}
\includegraphics[width=\linewidth]{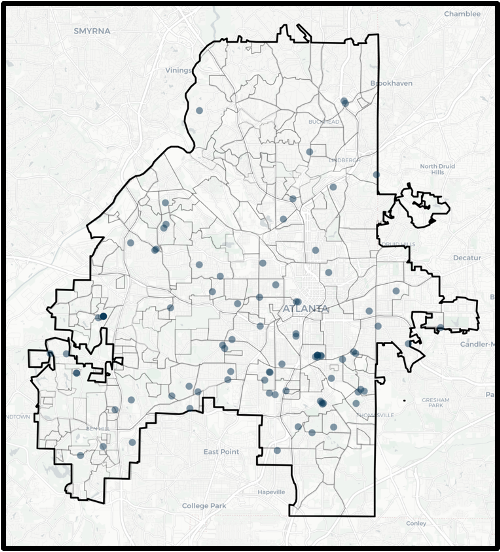}
\caption{July 4th, 2021}
\end{subfigure}
\begin{subfigure}[h]{0.4\linewidth}
\includegraphics[width=\linewidth]{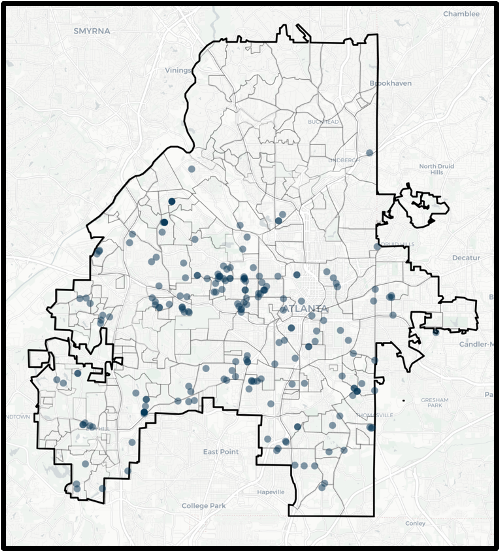}
\caption{December 31th, 2021}
\end{subfigure}
\caption{Snapshots of the spatial distributions of gunshot events on Independence Day and New Year's Eve in 2021, respectively.}
\label{fig:gunshot-snapshot}
\end{figure}

\paragraph{Why need spatio-temporal non-stationary models?}
In our scenario, point process models with a stationary influence kernel (\textit{e.g.}, ETAS) may lack the necessary flexibility. A stationary influence kernel is shift-invariant, meaning its value depends only on the difference between two time points or locations. Consequently, it cannot capture non-isotropic spatial influences or other complex spatial dependence structures. For example, gunshot events in Atlanta are heterogeneously distributed across different neighborhoods with a complex structure, as shown in Figure~\ref{fig:gunshot-distribution}. Moreover, the contagious effect of gunshot events is non-stationary and significantly influenced by the socioeconomic status and social network structure of local communities. This situation necessitates the use of a non-stationary model to effectively characterize the intricate patterns of gun violence in Atlanta.

\paragraph{Covariate data.}

\begin{figure}[!t]
\centering
\begin{subfigure}[h]{0.325\linewidth}
\includegraphics[width=\linewidth]{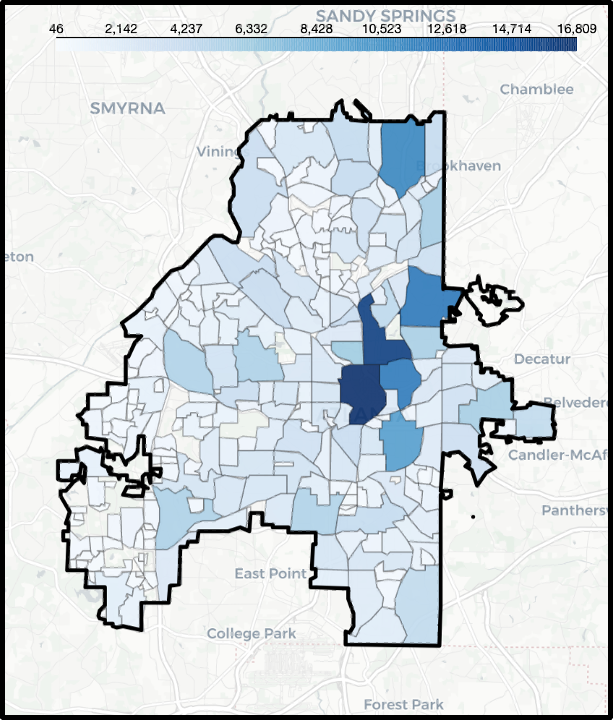}
\caption{Total population}
\end{subfigure}
\begin{subfigure}[h]{0.32\linewidth}
\includegraphics[width=\linewidth]{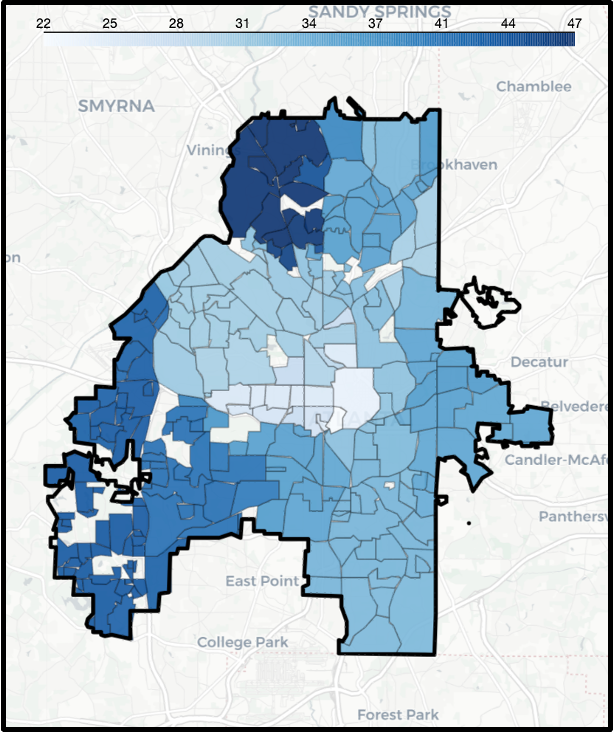}
\caption{Median age}
\end{subfigure}
\begin{subfigure}[h]{0.32\linewidth}
\includegraphics[width=\linewidth]{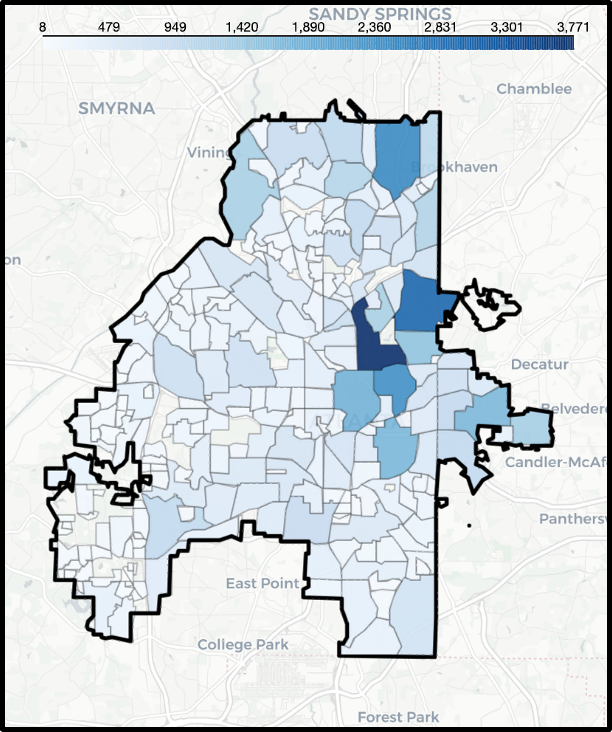}
\caption{Housing units}
\end{subfigure}

\begin{subfigure}[h]{0.325\linewidth}
\includegraphics[width=\linewidth]{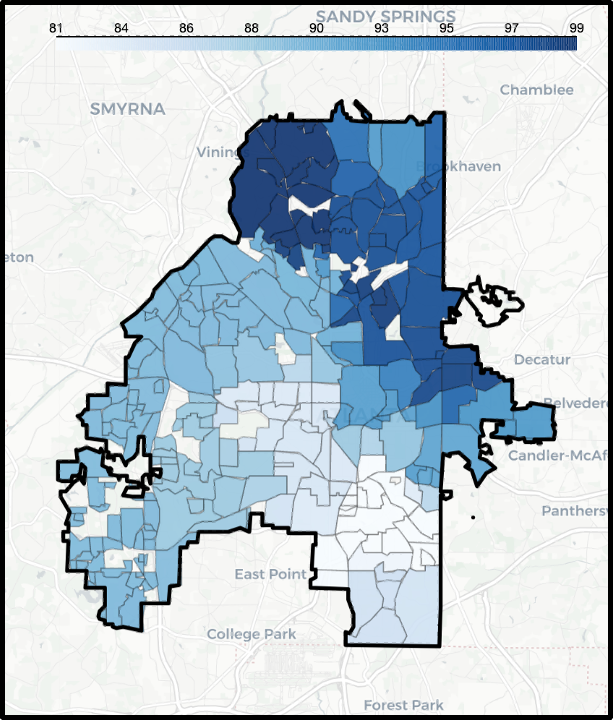}
\caption{High school percentage}
\end{subfigure}
\begin{subfigure}[h]{0.32\linewidth}
\includegraphics[width=\linewidth]{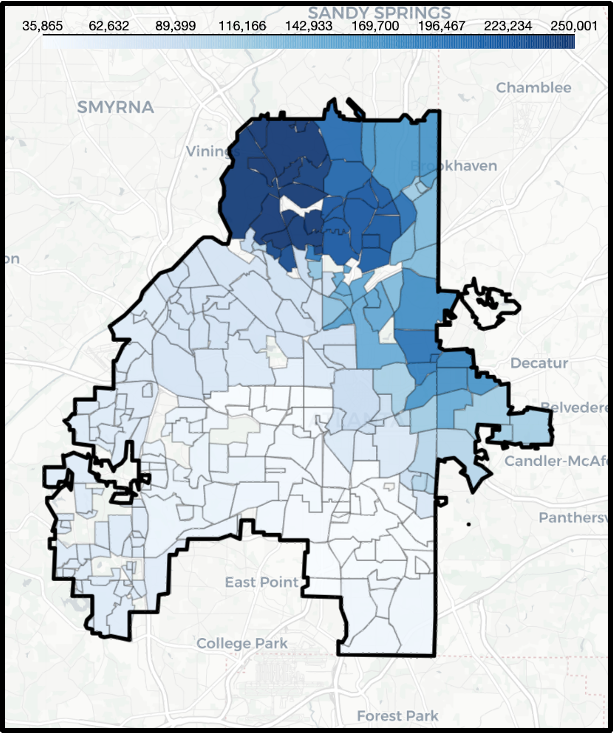}
\caption{Median income}
\end{subfigure}
\begin{subfigure}[h]{0.32\linewidth}
\includegraphics[width=\linewidth]{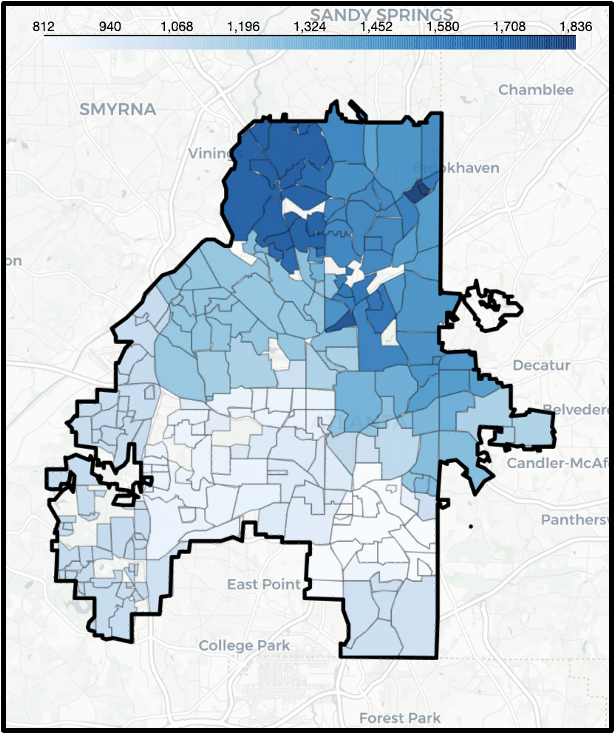}
\caption{Median rent}
\end{subfigure}
\caption{Six covariates were collected for each neighborhood to study their effect on local gunshot event intensity, using values reported in 2021. (a) Total population. (b) Median age of the population. (c) Occupied housing units. (d) Percentage of the population over 25 years old who have completed high school or higher degrees. (e) Median family income in the past 12 months. (f) Median rent of occupied rental units.}
\label{fig:demographic-variables}
\end{figure}

In addition to the gunshot data, publicly available census data provides crucial insights into the spatial covariate effects on the intensity of local gunshot events. This data is derived from the American Community Survey (ACS) 5-year dataset \citep{ACS5Y}, which includes over 66,000 variables covering a wide range of topics related to the socioeconomic characteristics of the U.S. population. We select six variables as our spatial covariates, representing demographic, economic, housing, and education figures, as shown in Figure~\ref{fig:demographic-variables}.
Analyzing the relationship between these covariates and gunshot events, we observe that areas with higher socioeconomic status, indicated by variables such as income levels and educational attainment, generally report fewer gunshot incidents. Conversely, regions with higher population density, such as urban centers, experience more frequent gunshot events. These observations suggest a potential relationship between the distribution of gunshot events and the socioeconomic characteristics of a neighborhood, as captured by our spatial covariates, supporting the assumptions of our model.

\section{Model}
\label{sec:model}

Motivated by the need to model spatio-temporal gunshot data with spatial covariates and to capture the non-stationary nature of gunshot spread, we adopt a point process model with a non-stationary kernel that incorporates spatial covariates for background intensity. In this section, we first revisit the background of self-exciting spatio-temporal point processes and then introduce our approach to modeling gun-related incidents in Atlanta using a spatial covariate-based point process model.

\subsection{Background}

Consider a sequence of $n$ incidents arranged in time order. Denote the $i$-th incident as a tuple $(t_i, s_i)$, where $t_i \in [0, T]$ represents the time of the occurrence of the incident, and the $s_i \in \mathcal{S}$ represents the location of the incident, measured in longitude and latitude coordinates; $\mathcal{S}$ denotes the observation domain, which for instance, can be the geographical area covered by the city of Atlanta.
The entire event sequence is given by
\begin{equation}
    (t_1, s_1), (t_2, s_2), \dots,(t_n, s_n).
    \label{eq:raw-data}
\end{equation}

Spatio-temporal point processes (STPPs) \citep{reinhart2018review} have been widely used to model sequences of discrete events that occur in continuous time and space. Let $\mathcal{H}_T = \{(t_i, s_i)\}_{i=1}^{n}$ denote the entire event sequence in $[0, T]$ and $\mathcal{H}_t = \{(t_i, s_i) \in \mathcal{H}|t_i < t\}$ denote the observed history before time $t$. 
An STPP is fully characterized by the \textit{conditional intensity function}
\[
    \lambda(t, s | \mathcal{H}_t) =  \lim_{\Delta t \downarrow 0, \Delta s \downarrow 0} \frac{\mathbb{E}\left[\mathbb{N}([t, t+\Delta t] \times B(s, \Delta s)) | \mathcal{H}_t\right]}{|B(s, \Delta s)| \Delta t},
\]
where $B(s, \Delta s)$ is a ball centered at $s \in \mathbb{R}^{2}$ with radius $\Delta s$, and the counting measure $\mathbb{N}$ is defined as the number of events occurring in $[t, t+\Delta t] \times B(s, \Delta s) \subset \mathbb{R}^{3}$. Therefore, the conditional intensity function can be interpreted as the rate of the occurrence of future events at time $t$ and location $s$. Naturally $\lambda\left(t, s | \mathcal{H}_t\right) \geq 0$ for any arbitrary $t$ and $s$. In the following, we omit the dependency on history $\mathcal{H}_t$ and use common shorthand $\lambda(t, s)$.

\vspace{.1in}
Hawkes processes \citep{hawkes1971spectra} is a type of self-exciting point process that captures the triggering effects between events. Assuming that influences from past events are linearly additive to the current event, the conditional intensity for a Hawkes point process takes the form of
\begin{equation}
    \lambda(t, s) = \mu(s) + \sum_{(t', s') \in \mathcal{H}_t}k(t, t', s, s'),
    \label{eq:pp-with-influence-kernel}
\end{equation}
where $\mu(s) > 0$ is a location dependent background rate of events at $s$, and $k(t, t', s, s')$ is the so-called \textit{influence kernel} that captures the influence of a past event happened at $(t', s')$ on the likelihood of event occurrence at a future time $t$ and location $s$.

\vspace{.1in}
Point processes can be represented using the conditional intensity function $\lambda$ from data, which is typically parameterized by model parameters denoted as $\theta$. Given the intensity function, the log-likelihood of the observed events $\mathcal{H}_T$ can be written as
\begin{equation}
    \ell(\theta|\mathcal{H}_T)=\sum_{i=1}^n \log \lambda(t_i, s_i)-\int_0^T \int_{\mathcal{S}} \lambda(t, s) ds dt.
    \label{eq:pp-log-likelihood}
\end{equation}
 The parameters can be estimated using maximum likelihood estimation (MLE) that solves $\max_\theta \ell(\theta|\mathcal{H}_T)$. 
To numerically solve the MLE problem, two common approaches include EM algorithm or gradient descent \citep{reinhart2018review}. We will adopt the latter approach and discuss further in Section \ref{sec:model-fitting}.

\subsection{Non-stationary kernel with deep neural networks}

We use a non-stationary kernel to model the contagion of gunshot events, which is further represented by neural networks. Our kernel allows for the modeling of a heterogeneous triggering effect of gunshot events in different regions. For model simplicity and computational efficiency, we adopt the commonly used assumption that the triggering effect of a past event is separable in space and time:
\[
    k(t, t^\prime, s, s^\prime) = \nu(t, t^\prime) \cdot \upsilon(s, s^\prime),
\]
where $\nu(t, t^\prime)$ and $\upsilon(s, s^\prime)$ are kernels that capture the temporal and spatial dependence between two events, respectively. 

\paragraph{Stationary temporal kernel.} 

As the goal focuses on capturing the non-homogenous spatial effect, we will assume a relatively simple temporal shift-invariant kernel with exponential decay:
\[
    \nu(t, t^\prime) =  C e^{-\frac{1}{2\sigma_0^2}(t-t^\prime)^2}, \quad t>t'.
\]
Here $C>0$ is a parameter that controls the magnitude of the kernel, and $\sigma_0 >0$ is a parameter that controls the decaying rate of the event's temporal influences. We assume $t>t'$ to capture the fact that a historical event at time $t'$ impacts the current time $t$ but not vice versa.

\paragraph{Non-stationary spatial kernel.} 
Given two arbitrary locations $s, s^\prime \in \mathcal S$, we define the spatial kernel $\upsilon(s, s^\prime)$ as a inner product between two feature mappings $\phi_s$ and $\phi_{s'}$, i.e,
\[
    \upsilon(s, s^\prime) = \left <\phi_s, \phi_{s^\prime} \right >, \quad s, s' \in \mathcal S,
\]
where the inner product for functions $\left <f, g \right > \coloneqq \int_{\mathbb{R}^2}f(u)g(u)du$. 
We represent the feature mapping
$\phi_s$ as a weighted sum of a set of $R$ independent kernel-induced feature functions $\{\kappa^{(r)}_s \coloneqq \kappa^{(r)}(s, \cdot)\}_{r=1}^{R}$:
\[
    \phi_s = \sum_{r=1}^{R}w_s^{(r)}\kappa^{(r)}_s,
\]
where 
$\kappa^{(r)}: \mathcal{S} \times \mathcal{S} \rightarrow \mathbb{R}_+$ is a general kernel and
$w_s^{(r)}$ is the corresponding weight of that feature function at location $s$. The location-dependent weight satisfies $\sum_{r=1}^{R}w_s^{(r)} = 1$ at any arbitrary location $s$. 
Hence the spatial kernel can be re-written as
\[
    \upsilon(s, s^\prime) = \sum_{1 \leq r_1, r_2 \leq R} w_s^{(r_1)}w_{s^\prime}^{(r_2)} \left < \kappa_s^{(r_1)}, \kappa_{s^\prime}^{(r_2)} \right >.
\]

%
The rationale of this design is two-fold: 
(a) Using a linear combination of the product of feature functions enhances the representative power of the spatial kernel. Note that when $r_1 = r_2$, the kernel captures self-similarity of feature functions and otherwise captures the cross-similarity between two feature functions. (b) The spatial kernel can also be highly interpretable if $\kappa^{(r)}$ takes a specific parametric form; following the idea in \cite{Higdon1998NonStationarySM, zhu2021early}, we choose $\kappa_s$ to be a Gaussian function centered at $s$ with covariance matrix $\Sigma_s$ since the spatial correlation between two events decays as their distance increases in general. The spatial kernel is specified to be:
\begin{equation}
    \upsilon(s, s^\prime) = \sum_{1\le r_1, r_2 \le R} \frac{w_s^{(r_1)}w_{s^\prime}^{(r_2)}}{2\pi | \Sigma_s^{(r_1)} + \Sigma_{s^\prime}^{(r_2)} |^{\frac{1}{2}}} \exp\left \{-\frac{1}{2}(s - s^\prime)^\top(\Sigma_s^{(r_1)} + \Sigma_{s^\prime}^{(r_2)})^{-1}(s - s^\prime)\right \}.
    \label{eq:spatial-kernel-expression-one-component}
\end{equation}

Now we specify the kernel-induced feature function $\kappa_s$.
According to \cite{Higdon1998NonStationarySM}, there exists a one-to-one mapping between a bivariate normal distribution specified by $\Sigma_s$ and its one standard deviation ellipse. 
Note that $\kappa_s$ is centered at $s$, so the ellipse's center is fixed at $s$. 
Thus we can specify the ellipse by a pair of focus points and its area $A$. The focus points are denoted by $\bm{\psi}_s = (\bm{\psi}_x(s), \bm{\psi}_y(s))$ and $-\bm{\psi}_s = (-\bm{\psi}_x(s), -\bm{\psi}_y(s))$, where $\bm{\psi}_s \in \Psi \subset \mathbb{R}^2$. 
Hence, given $\bm{\psi}_s$ and $A$, the corresponding $\Sigma_s$ can be written as
\[
    \Sigma_s = \tau_z^2 \begin{pmatrix} Q + \frac{\|\bm{\psi}_s\|^2}{2}\cos2\alpha & \frac{\|\bm{\psi}_s\|^2}{2}\sin2\alpha \\ \frac{\|\bm{\psi}_s\|^2}{2}\sin2\alpha & Q - \frac{\|\bm{\psi}_s\|^2}{2}\cos2\alpha \end{pmatrix},
\]
where $Q = \sqrt{4A^2 + \|\bm{\psi}_s\|^4\pi^2}/2\pi, \alpha = \tan^{-1}(\bm{\psi}_y(s) / \bm{\psi}_x(s))$, $\tau_z > 0$ is a scaling parameter that controls the overall level of the covariance. For computational simplicity, we assume the area of the ellipse remains to be $A$ over all locations, and treat $A$ as a hyperparameter.


\paragraph{Neural network-based kernel representation.}

We develop a neural network-based representation for the kernel-induced feature function similar to the idea in \cite{zhu2021early, zhu2022neural}. A key feature of our non-stationary spatial kernel is that for any location $s \in \mathcal{S}$, we estimate a mapping that obtains the focus point $\bm{\psi}_s$ and the corresponding location-dependent weight $w_s$. To this end, we represent the mapping $\varphi: \mathcal{S} \rightarrow \Psi \times [0, 1]$ from the location to the space of focus points $\Psi$ and the weights $[0, 1]$ using a fully-connected multi-layer neural network. The input of the neural network is the two-dimensional location vector $s$, and the output is the concatenation of the corresponding focus point $\bm{\psi}_s$ and its weight $w_s$. In our implementation, we adopt the same network architecture for all $R$ kernel-induced feature functions. Each of them is a fully-connected network with two hidden layers, and each hidden layer is equipped with the Softplus activating function $f(x) = \log(1+e^x)$. We will numerically show that in our setting, neural networks provide a flexible representation of the covariance and the corresponding kernel-induced feature function due to their well-known universal approximation power. 


\subsection{Modeling spatial covariates}

Spatial covariates, such as demographic variables, also contribute to the spatial heterogeneity of gunshot occurrences. To characterize the exogenous effect of spatial covariates on the gunshot event intensity over space, we first discretize the entire observation domain $\mathcal{S}$ into multiple subregions $\{\mathcal{S}_1, \mathcal{S}_2, \dots, \mathcal{S}_J\}$, and then collect $L$ spatial covariates for each subregion $\mathcal{S}_j$, denoted by $\{x_{j1}, \dots, x_{jL}\}$, based on which we can establish the connection between the spatial covariates and the gunshot event intensity.

\vspace{.1in}
Specifically, we incorporate the covariate effect by introducing a location-dependent background intensity in the conditional intensity function:
\begin{equation}
    \mu(s) = \mu_0 + \exp\left \{\sum_{l=1}^{L}\sum_{j \in N(s)}\gamma_l x_{jl} \cdot \left( \frac{e^{-\alpha d(s, c_j)}}{\sum_{j'\in N(s)} e^{-\alpha d(s, c_{j'})}} \right)\right \}.
    \label{eq:census-exogenous-bg-intensity}
\end{equation}
Here, we assume that the background intensity $\mu(s)$ is time-invariant and use a weighted sum of the spatial covariates to characterize their exogenous effect on the gunshot event intensity. The bandwidth parameter $\alpha > 0$ controls the magnitude of the covariate effect. The $c_j$ represents the geographic centroid of the subregion $\mathcal{S}_j$, and $d(s, c_j)$ measures the Euclidean distance between location $s$ and $c_j$. We only consider the effects of covariates in the subregion that the location $s$ falls in and other subregions that are geographically adjacent to it, by summing over the corresponding index set $N(s)$.

\vspace{.1in}
Each covariate may positively or negatively contribute with event intensity. Thus, we introduce a coefficient $\gamma_l$ for the $l$-th covariate to indicate such a correlation and its magnitude, which is learned from data. The exponential function outside the weighted sum ensures the non-negativity of the intensity, and $\mu_0 > 0$ captures the endogenous intensity of gunshots.  
The weighted sum of the spatial covariates allows for smooth modeling of the exogenous covariate effect on the gunshot events. 



\section{Model fitting}
\label{sec:model-fitting}

The learning of our model requires the estimation of parameters in deep neural networks. We solve the optimization problem of maximum likelihood estimation (MLE) using gradient descent (GD), which has proven to be effective for model estimation when a large number of parameters is involved \citep{reinhart2018review}. The learning rate is set to be $0.1$, and the learning process continues until convergence.

\vspace{.1in}
In numerically evaluating the log-likelihood function \eqref{eq:pp-log-likelihood}, which is the objective function of the optimization problem, two major difficulties arise. These challenges are also typical in model fitting for spatio-temporal point process models using maximum likelihood, particularly in dealing with the double integral of the conditional intensity function in \eqref{eq:pp-log-likelihood}. Here, the double integral can be re-written as
\begin{equation}
    \begin{aligned}
        \int_0^T \int_{\mathcal{S}} \lambda(t, s) ds dt &= \mu_0|\mathcal{S}|T + \int_0^T \int_{\mathcal{S}} \sum_{(t', s') \in \mathcal{H}_t} k(t, t', s, s') ds dt \\&+ \int_0^T \int_{\mathcal{S}} \exp\left \{\sum_{l=1}^{L}\sum_{j \in N(s)}\gamma_l x_{jl} \cdot \left( \frac{e^{-\alpha d(s, c_j)}}{\sum_{j'\in N(s)} e^{-\alpha d(s, c_{j'})}} \right)\right \} ds dt.
    \end{aligned}
    \label{eq:our-model-loglikelihood}
\end{equation}

\vspace{.1in}
The first challenge arises from the integral of the influence kernel (the second term) involves the spatio-temporal integral of the summation of the effects from historical events, leading to a computational complexity of $\mathcal{O}(n^3)$ with a total of $n$ events in the event series. To reduce the computational cost, we adopt an approximation strategy similar to that used in \citep{dong2023non}--- computing the integral of influence kernel with a complexity of only $\mathcal{O}(n)$. Specifically, we assume that the distance between two focus points of an ellipse ($\bm\psi_s$ and $-\bm\psi_s$) at an arbitrary location $s$ is bounded by a threshold $2c$, for a large distance between focus points can lead to an overstretched ellipse, which is unrealistic in practice. Then, by approximating the kernel-induced feature function $\kappa_s$ using a Gaussian function with an isotropic one-standard-deviation ellipse centered at $s$ of area $A$, the integral of the influence kernel can be approximated by
\[
    \int_{0}^{T}\int_{\mathcal{S}} \sum_{(t', s') \in \mathcal{H}_t} k(t, t', s, s')dtds = (1 + \epsilon) \left\{ \sqrt{2\pi}C\sigma_0 \sum_{i=1}^{n}\left[h\left(\frac{T - t_i}{\sigma_0}\right) - \frac{1}{2} \right] \right \},
\]
where $\epsilon$ represents the relative error of the integral approximation, and $h$ is the density function of standard Gaussian distribution. The $C$ and $\sigma_0$ are the parameters in the temporal kernel.
We note that the relative error is bounded by $|\epsilon| < \max \left\{ U-1, 1- 1/U \right\}$, where $U = (\sqrt{4A^2 + c^4\pi^2} + c^2 \pi)/2A$. In our experiments, we choose $A = 0.35, c = 0.1$ to control the relative error under $0.05$.

\vspace{0.1in}
Another challenge arises from the integral of the covariate effect. We achieve efficient computation of this integral through a similarly tailored numerical integration strategy. Denoting the entire exponential function as $g(s)$, we overlay a sufficiently dense grid $\mathcal{U} \in \mathcal{S}$ over the space. The function $g(s)$ is then evaluated at each grid point in $\mathcal{U}$, allowing the integral to be approximated by
\[
    \begin{aligned}
        \int_0^T \int_{\mathcal{S}} \exp\left \{\sum_{l=1}^{L}\sum_{j \in N(s)}\gamma_l x_{jl} \cdot \left( \frac{e^{-\alpha d(s, c_j)}}{\sum_{j'\in N(s)} e^{-\alpha d(s, c_{j'})}} \right)\right \} ds dt &\triangleq \int_{0}^{T}\int_{\mathcal{S}}g(s) dsdt \approx T\frac{|\mathcal{S}|}{|\mathcal{U}|}\sum_{u \in \mathcal{U}}g(u).
    \end{aligned}
\]
Note that the $\{N(u), \{c_j, d(u, c_j)\}_{j \in N(u)}\}_{u \in \mathcal{U}}$ can be pre-computed once the grid $\mathcal{U}$ is given. Therefore, we only need to plug in the values of $\{\gamma_l\}_{l \in L}$ every time the parameters are updated during the GD process.

\vspace{0.1in}
When fitting the model with spatial covariates $x_{jl}$, we standardize the data by subtracting the mean from each covariate and dividing by its respective standard deviation. In our dataset, all spatial covariates are numerical. This standardization ensures that each covariate is ``comparable'' in magnitude.

\section{Results}
\label{sec:results}

Below, we present the numerical results of fitting our model to the Atlanta gunshot and census data. Our analysis covers several aspects: in-sample estimation to validate the model's fit to the data, out-of-sample predictions to assess its forecasting accuracy, and visualization of the learned kernels to highlight the model's interpretability.

\subsection{In-sample estimation}

\begin{figure}
\centering
\begin{subfigure}[h]{0.63\linewidth}
\includegraphics[width=\linewidth]{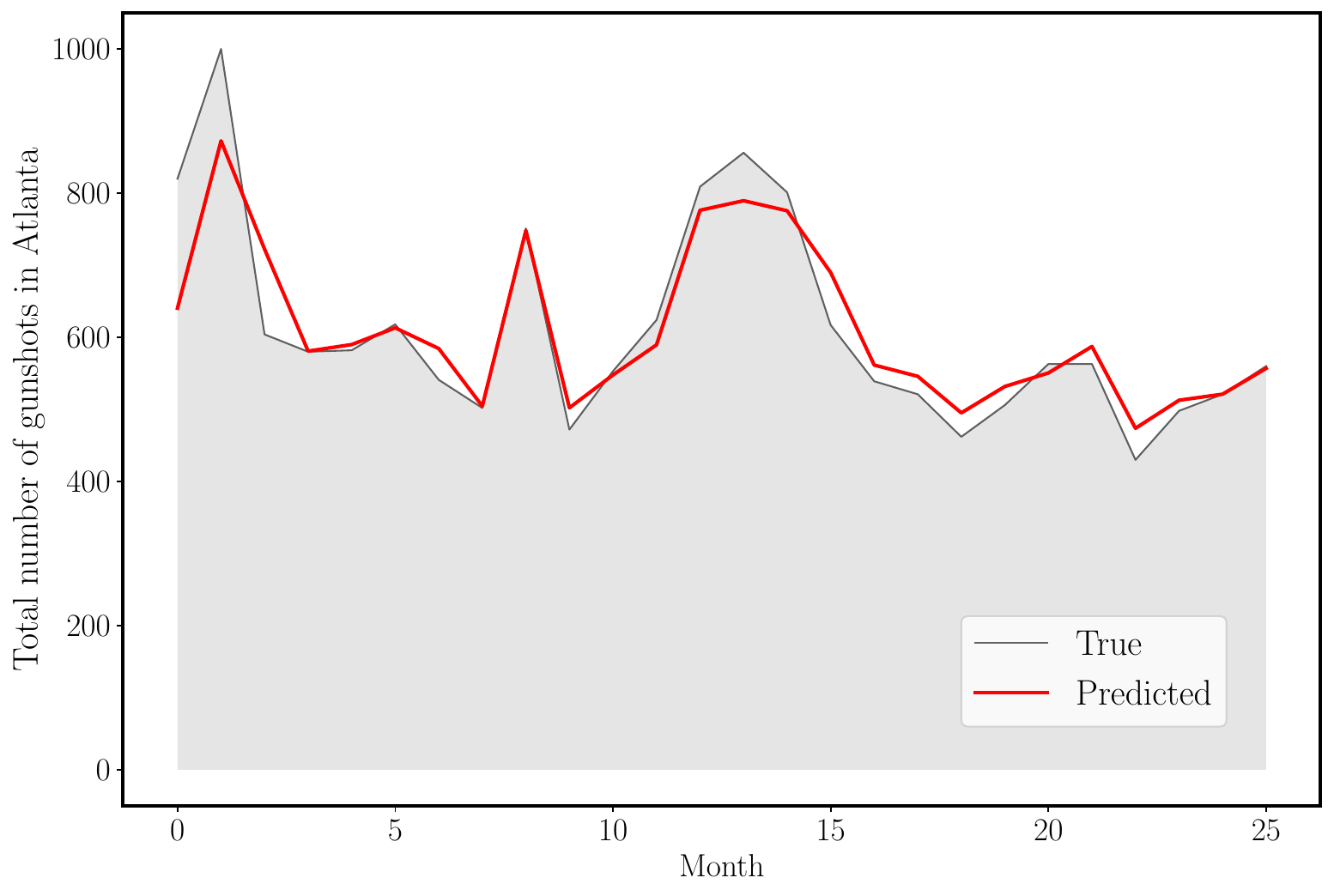}
\end{subfigure}
\begin{subfigure}[h]{0.36\linewidth}
\includegraphics[width=\linewidth]{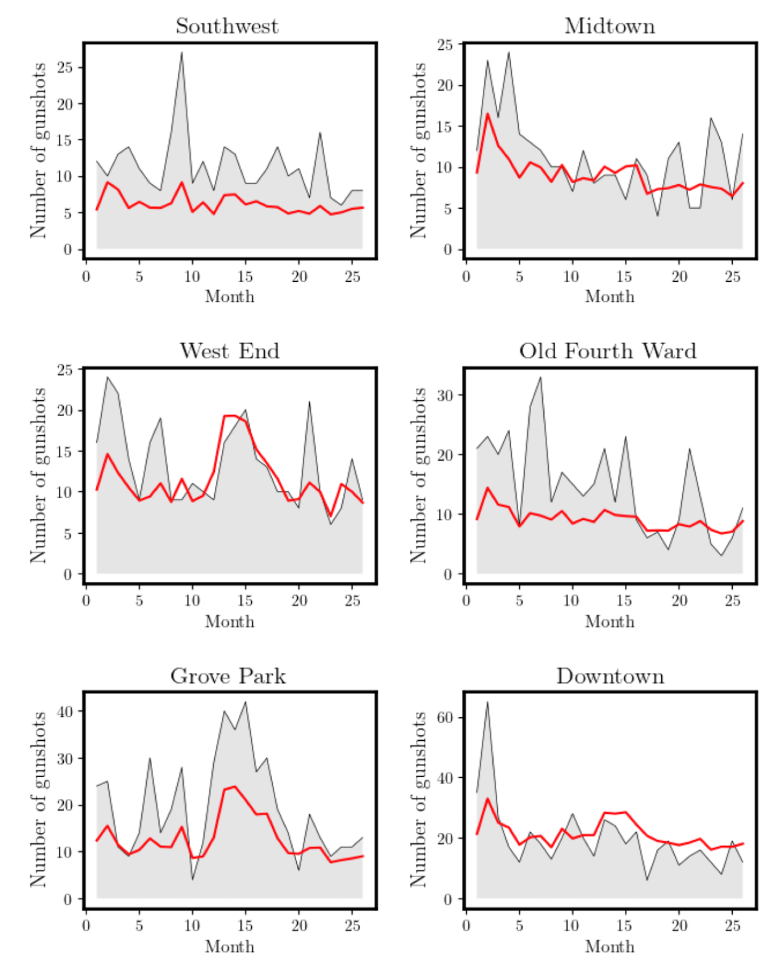}
\end{subfigure}
\caption{In-sample estimations of the monthly number of gunshot events from May 11, 2021, to May 10, 2023. The red lines represent the model estimations, and the black lines represent the true observations. Left panel: monthly estimation of the total number of gunshots in the city of Atlanta. Right six panels: monthly estimations in the top six neighborhoods with the highest number of gunshots. The mean absolute error (mean relative error) of the estimations for total event numbers is 36.919 (5.64\%).}
\label{fig:in-sample-estimation}
\end{figure}

We fit the model to two-year gunshot data and analyze its performance. To demonstrate the effectiveness of our model, we perform in-sample predictions for the number of gunshot events in different regions. The in-sample prediction is conducted as follows: given a time interval denoted as $[t_1, t_2]$, we input the training data into the fitted model and calculate the integral of the conditional intensity function over $[t_1, t_2]$ to estimate the number of gunshot events within that time interval. In our evaluation, we estimate the monthly event numbers using our model, including both the total number of gunshot events within the city of Atlanta and the number of events in each neighborhood. 

Figure~\ref{fig:in-sample-estimation} visualizes the in-sample estimations produced by our model and compares them with the actual number of reported gunshot events. Our model excels in accurately estimating the number of gunshot events at both the city and neighborhood levels, indicating a good fit to the data. Despite variations in event numbers across different times and neighborhoods, our model effectively captures the spatio-temporal dynamics. We also report performance metrics, including the mean absolute error (MAE) and mean relative error (MRE), for the in-sample estimation of the total event numbers in Atlanta, which indicate reasonably good performance, with a relative error of 5.64\%.

\subsection{Interpretable spatial kernel}

\begin{figure}[!t]
\centering
\begin{subfigure}[h]{0.32\linewidth}
\includegraphics[width=\linewidth]{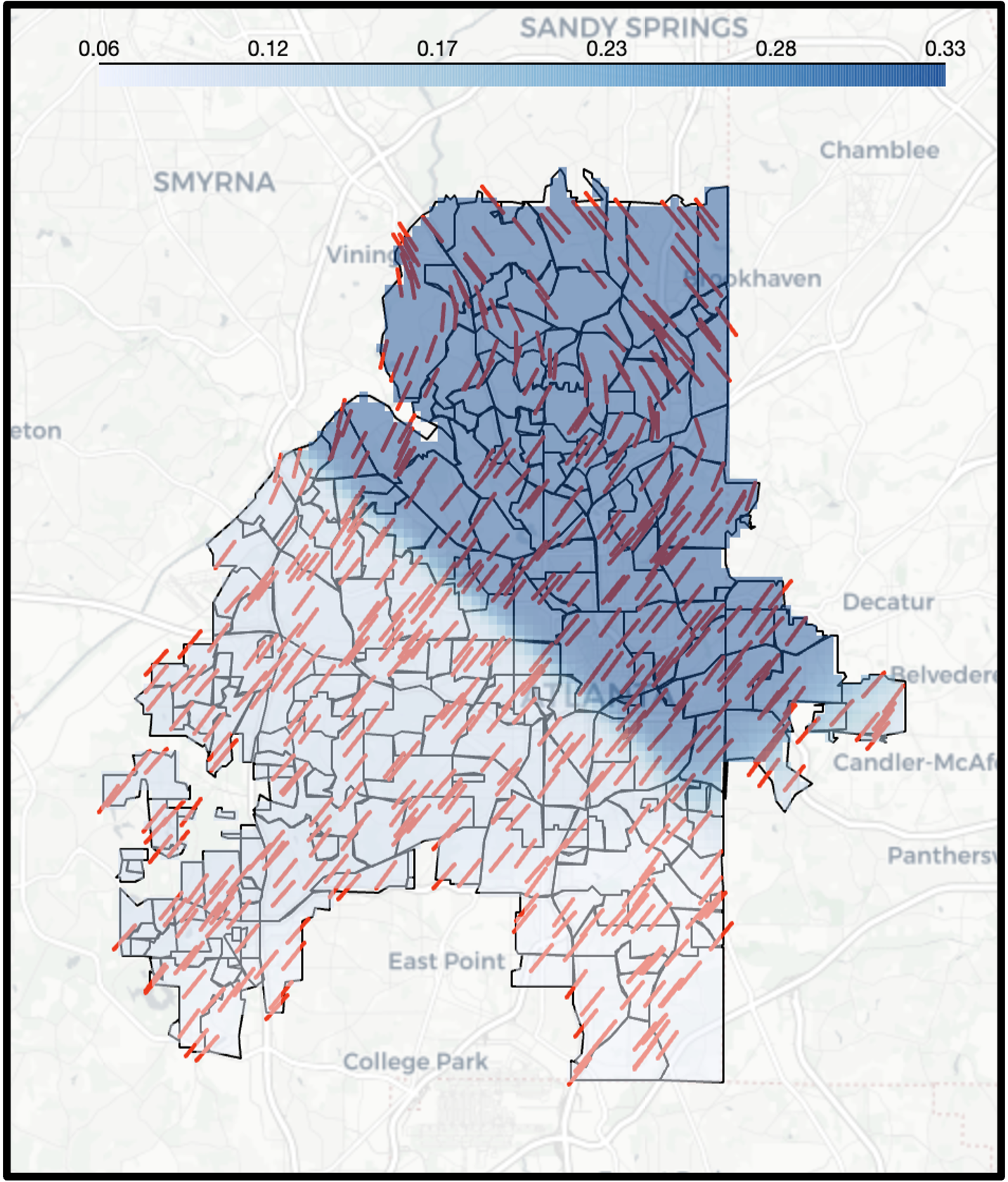}
\caption{$\kappa_s^{(1)}$}
\end{subfigure}
\begin{subfigure}[h]{0.32\linewidth}
\includegraphics[width=\linewidth]{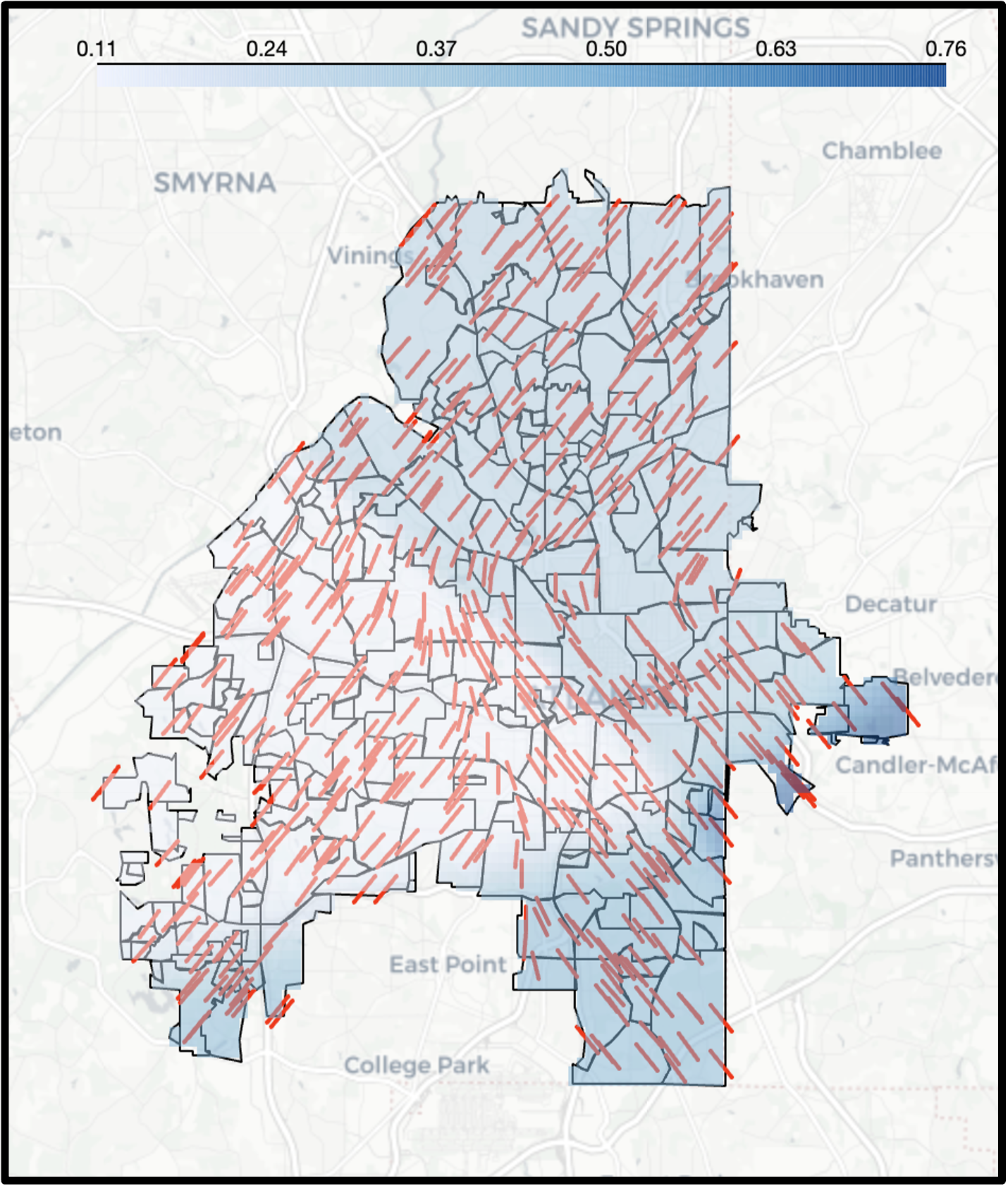}
\caption{$\kappa_s^{(2)}$}
\end{subfigure}
\begin{subfigure}[h]{0.32\linewidth}
\includegraphics[width=\linewidth]{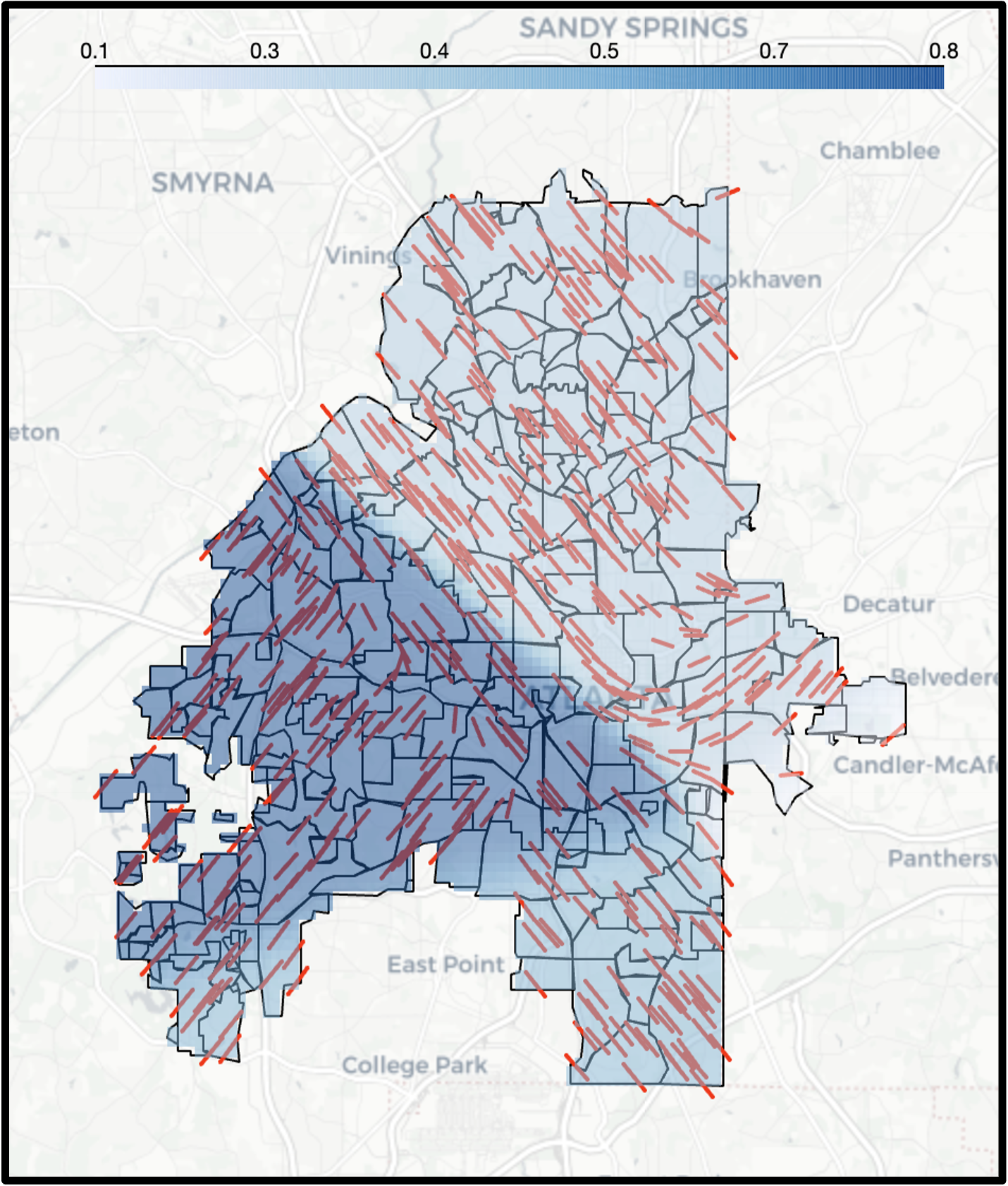}
\caption{$\kappa_s^{(3)}$}
\end{subfigure}
\caption{Three distinctive modes of the spatial kernel for gunshot events are learned from the data, as indicated by the three learned feature functions $\kappa_s^{(1)}, \kappa_s^{(2)}$, and $\kappa_s^{(3)}$. In this visualization, each red line segment connects two focus points of the ellipse corresponding to the Gaussian function $\kappa_s^{(r)}$ at location $s$ (200 line segments are sampled over space in each panel). The shaded area in blue shows the magnitude of the weight $w_s^{(r)}$ over space, with darker colors indicating larger weights.}
\label{fig:spatial-kernel-components}
\end{figure}

Our model also demonstrates strong explanatory power through the learned non-stationary spatial kernel. Recall that each $\kappa_s^{(r)}$ is a Gaussian function centered at $s$ with a spatially varying covariance matrix. By interpreting the covariance matrix, we can discover to some extent the underlying spatio-temporal dynamics of the contagious effect of gunshot events. In Figure~\ref{fig:spatial-kernel-components}, we present three learned feature functions that compose our spatial kernel by visualizing the covariance matrix at different locations. Specifically, we sample 200 locations across the space and draw a line segment at each location, connecting the two focal points of the ellipse corresponding to the covariance matrix. The angle and length of each red line can be interpreted as the direction and strength of the gunshot contagious effect at that specific location.

\vspace{.1in}
We observe three distinct modes of the gunshot contagious effect, as indicated by each feature function. The spatial heterogeneity of this effect is clearly evident, underscoring the necessity of our non-stationary kernel. Moreover, detailed characteristics of the gunshot contagious effect in Atlanta can be discerned. For instance, gunshot events in the southwestern area tend to have a broader impact on nearby neighborhoods compared to those in the northeastern part of Atlanta, providing valuable insights for local practitioners to focus more attention on gunshot risk in those areas.


\subsection{Estimated covariate effect}

\begin{table}[!t]
  \caption{Estimated $\{\gamma_l\}_{l=1}^{L}$ of spatial covariates.}
  \centering
  \resizebox{1.\linewidth}{!}{
  \begin{threeparttable}
    \begin{tabular}{p{4cm}p{3cm}|p{4cm}p{3cm}}
    \toprule
    \toprule
    Covariate $l$ & $\gamma_l$ & Covariate $l$ & $\gamma_l$ \\
    \midrule
    Total Population & $2.2851$ & High school percentage & $-1.8666$ \\
    Median age & $-1.9211$ & Median income & $-3.8588$ \\
    Housing units  & $-0.5737$ & Median rent & $-0.3127$ \\
    \bottomrule
    \bottomrule
  \end{tabular}
  \end{threeparttable}
  }
  \label{tab:estimated-covariate-effect}
\end{table}

Another feature of our model is its capability to estimate the covariate effects, which can be analyzed through the fitted coefficients $\{\gamma_l\}_{l=1}^{L}$. Table~\ref{tab:estimated-covariate-effect} presents the fitted $\{\gamma_l\}_{l=1}^{L}$ alongside the corresponding covariate names. 
The sign of each coefficient provides crucial insights into the interpretation of the covariate effect. For example, while a larger population size in a neighborhood tends to increase the likelihood of observing gunshot events, other covariates, such as the high school graduation rate and median income, exhibit a negative (inhibitive) effect on gunshot intensity. These inhibitive covariate effects align with the understanding that gun violence is less prevalent among wealthier and more educated populations.
The magnitude of the coefficients also highlights the importance of considering these covariates when evaluating gun violence risk. The results indicate that the covariates of total population and median income have the largest effect magnitudes. We would like to emphasize that we can compare the magnitude of the coefficients, because we standardize the covariates before being input into the model.

\subsection{Out-of-sample prediction}

\begin{figure}[!t]
\centering
\begin{subfigure}[h]{0.32\linewidth}
\includegraphics[width=\linewidth]{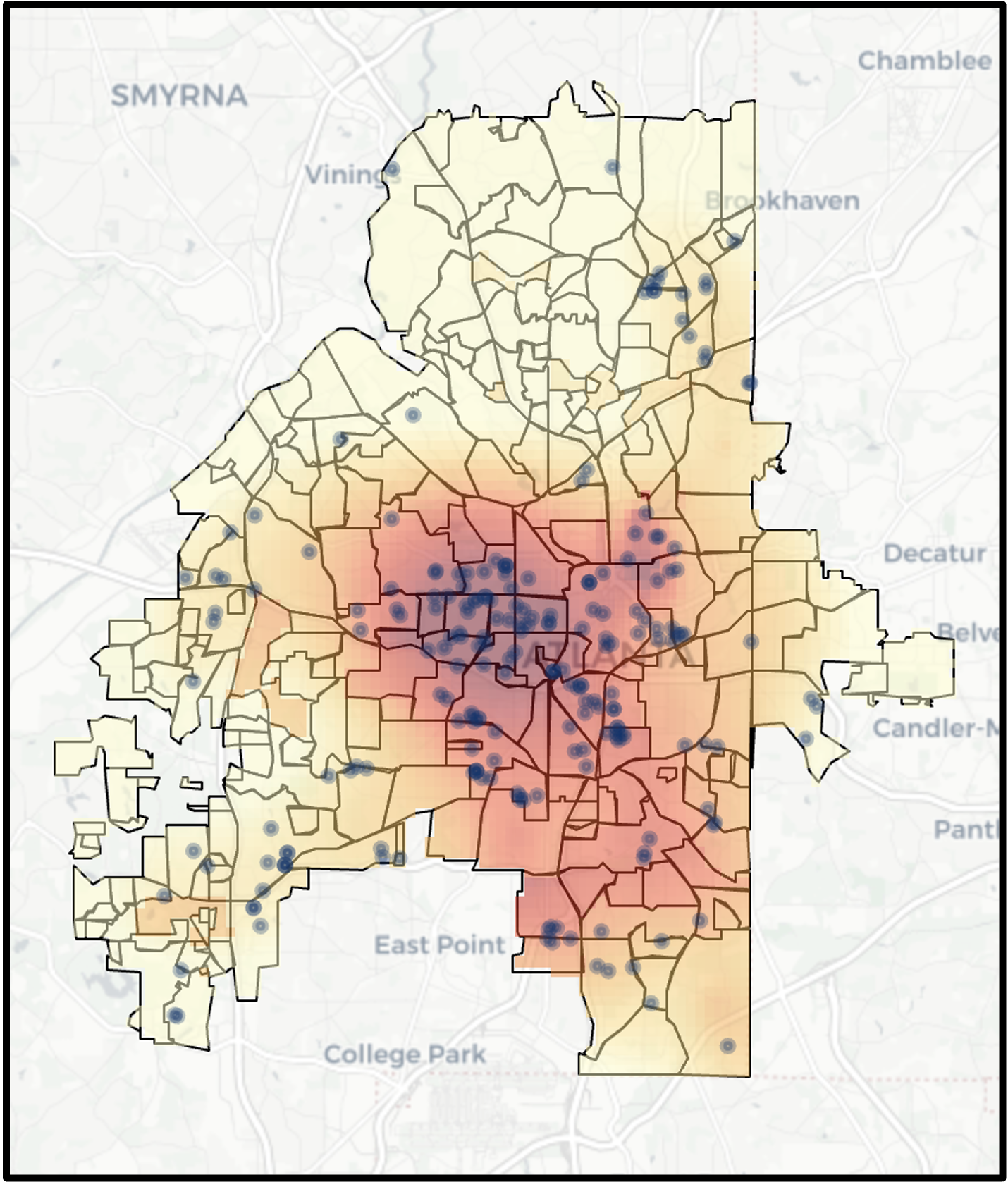}
\caption{May 22, 2023}
\end{subfigure}
\begin{subfigure}[h]{0.32\linewidth}
\includegraphics[width=\linewidth]{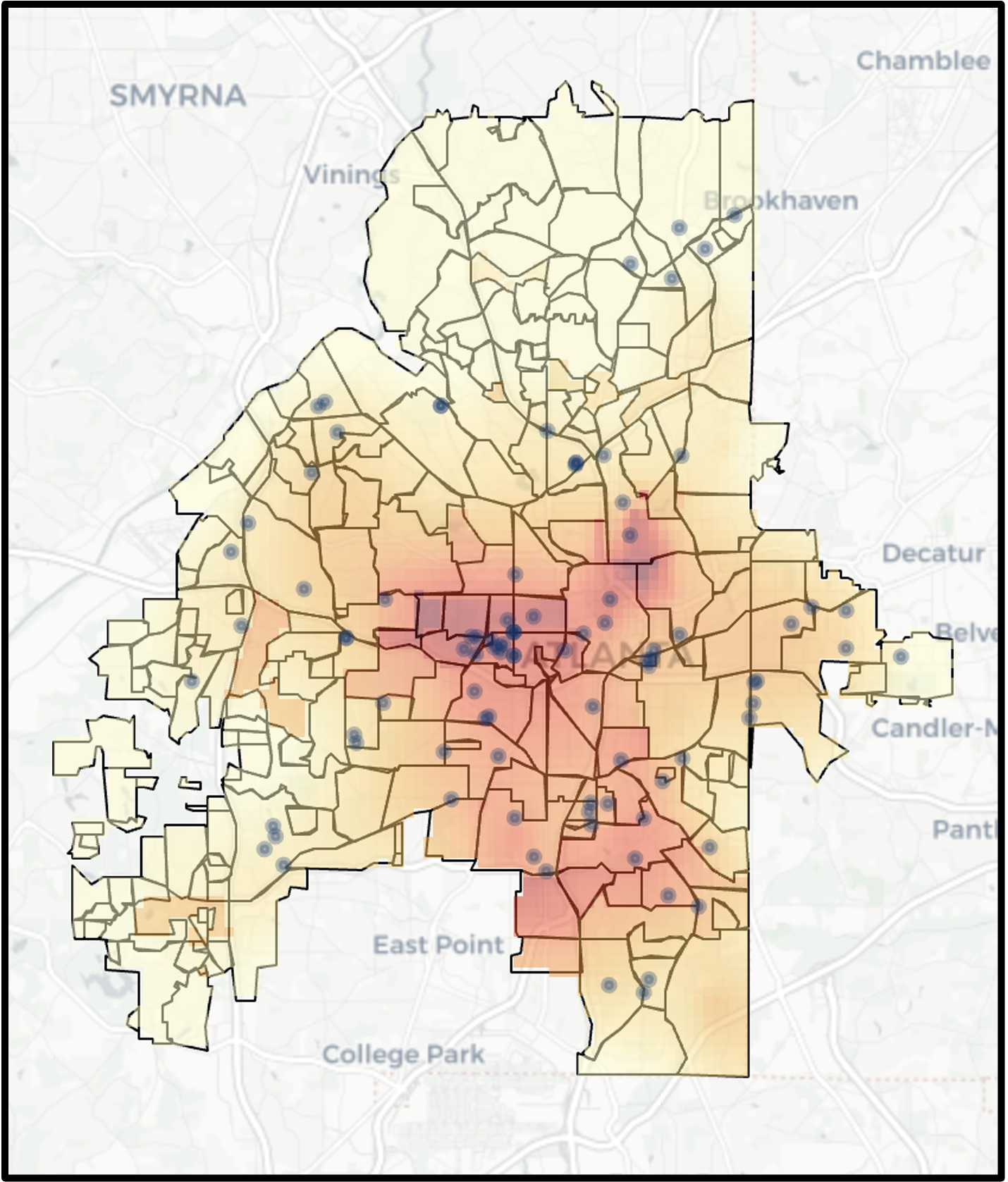}
\caption{June 09, 2023}
\end{subfigure}
\begin{subfigure}[h]{0.32\linewidth}
\includegraphics[width=\linewidth]{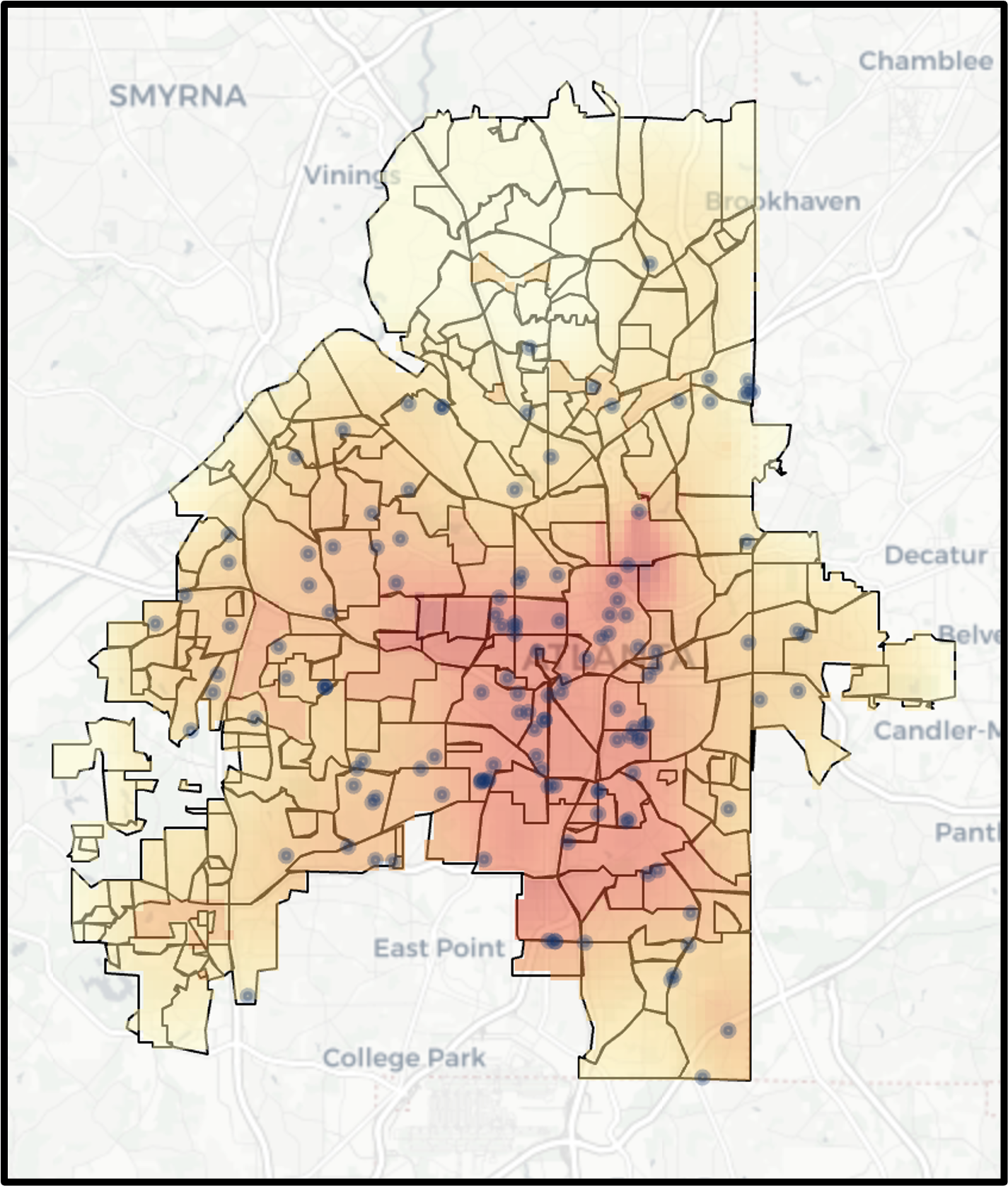}
\caption{June 20, 2023}
\end{subfigure}
\caption{Predicted conditional intensity function over space at three testing weeks. The color depth represents the magnitude of the predicted intensity. The dots are reported gunshot events during that week. Note that these events are testing data that are not included for predicting the conditional intensity, and our predictions align well with these events.}
\label{fig:oos-prediction-over-space}
\end{figure}

To assess the model's ability to forecast future events, we use the fitted model to perform out-of-sample predictions for the conditional intensity function from May 11, 2023, to June 24, 2023, and compare these predictions with the actual gunshot data collected during this period.
In out-of-sample prediction, the predicted conditional intensity function can be used to forecast gunshot hotspots. The conditional intensity function at any given time $t^*$ during the testing period can be evaluated by inputting the data prior to $t^*$ into the learned model.
Figure~\ref{fig:oos-prediction-over-space} displays the predicted conditional intensity over space at three different weeks during the testing period. We observe that the predicted intensity from our model aligns well with the spatial distribution of gunshot occurrences, with higher values predicted in areas that subsequently experienced more gunshot events in the following days.

\begin{table}[!t]
  \caption{Out-of-sample prediction performance by different models.}
  \centering
  \resizebox{0.85\linewidth}{!}{
  \begin{threeparttable}
    \begin{tabular}{p{4cm}>{\centering}p{3cm}>{\centering}p{3cm}>{\centering\arraybackslash}p{3cm}}
    \toprule
    \toprule
    Model & MAE (rare) & MAE (frequent) & MAE (total) \\
    \midrule
    \texttt{Persistent} & $0.337$ & $2.116$ & $0.679$ \\
    \texttt{ARIMA} & $0.354$ & $1.763$ & $0.625$ \\
    \texttt{ETAS} & $0.392_{(0.002)}$ & $2.827_{(0.024)}$ & $0.792_{(0.008)}$ \\
    Our model & $\textbf{0.305}_{(0.003)}$ & $\textbf{1.694}_{(0.018)}$ & $\textbf{0.572}_{(0.006)}$ \\
    \bottomrule
    \bottomrule
  \end{tabular}
  \end{threeparttable}
  }
  \label{tab:oos-baseline-comparison}
\end{table}

\vspace{.1in}
A more quantitative way to assess the model's predictive ability is to compute the predicted number of gunshot events via the conditional intensity function and evaluate the prediction error using actual observations. We perform out-of-sample predictions for the number of gunshot events on a weekly basis. For a given time $t^*$, the predicted event number in a neighborhood during the following week is obtained by integrating the conditional intensity function over time and the neighborhood, similar to the in-sample estimations. The MAE between the predictions and the true observations is then computed to indicate the model's predictive performance. We also perform out-of-sample predictions using three baseline models over the same period and compare their performance with our model: (a) a standard persistent prediction model (\texttt{Persistent}) that uses the event number from the previous week as the prediction for the next week; (b) an ARIMA time series model (\texttt{ARIMA}); and (c) an Epidemic-type Aftershock Sequence (\texttt{ETAS}) model \citep{Ogata1988} with a stationary diffusion kernel in Euclidean space.


\vspace{.1in}
Table~\ref{tab:oos-baseline-comparison} reports three types of MAEs as the final evaluation metrics: MAE (frequent) represents the errors of event number predictions averaged over the top 20\% of neighborhoods with the highest number of gunshot events, MAE (rare) is the average prediction error over the remaining 80\% of neighborhoods, and MAE (total) is the average across all neighborhoods. Our model consistently achieves the lowest MAE across all three metrics, demonstrating superior performance compared to other baselines in forecasting gunshot events across neighborhoods with varying characteristics. While \texttt{ARIMA} exhibits performance metrics closest to our model, it is a time series model designed for predicting event numbers and is not intended for handling discrete spatio-temporal events or providing insights into the underlying event dynamics. 

\section{Discussions}\label{discussion}

We have presented a study using a novel point process model with a non-stationary spatial kernel to analyze gunshot event data in Atlanta. The non-stationary spatial kernel enables a more accurate representation of the complex contagious effects of gunshot events across space, as traditional models often fail to capture these dynamics with stationary spatial interactions. A heterogeneous background intensity that incorporates the effects of census factors accounts for the inherent variability in gunshot events across different communities. By integrating these elements, our model demonstrates superior performance compared to baselines, such as persistent prediction approaches and ETAS models with stationary kernels, in predicting the number and distribution of future gunshot events. The ability to outperform these baselines confirms the presence of intricate underlying dynamics in gunshot event data that are neither random nor simple repetitions of previous patterns. The performance gain of our model over \texttt{ETAS} (which uses a homogeneous kernel) underscores the necessity of adopting a non-stationary spatial kernel to capture the heterogeneous contagious effects of gunshot events across space.

\vspace{.1in}
Looking ahead, several avenues for further research are worth exploring: Developing a statistical framework to quantify the significance of the covariate effects (e.g., p-values) on gunshot events, which has yet to be developed. Such significance levels are crucial for confirming the predictive validity of the covariates and for informing policy formulation. Additionally, incorporating a non-stationary temporal kernel could help capture broader temporal trends and shifts in gunshot event patterns over time, potentially related to holiday effects or changes in law enforcement tactics. Furthermore, simulation tools for point process models could be employed to distinguish the heterogeneity of events arising from covariate effects versus contagious phenomena. Such differentiation could aid in developing more effective policies to prevent gun-related incidents.

\section*{Acknowledgement}

This work is partially supported by an NSF CAREER CCF-1650913, NSF DMS-2134037, CMMI-2015787, CMMI-2112533, DMS-1938106, DMS-1830210, the Atlanta Police Foundation, and the Coca-Cola Foundation. We would like to thank Giuliano Tissot for the helpful discussions.

\bibliographystyle{apalike}
\bibliography{refs}

\end{document}